\documentclass[showpacs,twocolumn,aps,floatfix,superscriptaddress,noshowpacs]{revtex4}
\usepackage[mathlines]{lineno}

\usepackage{dsfont}
\usepackage{amsmath,amssymb,graphicx,bm,color,mathrsfs,verbatim,epstopdf,dcolumn,cancel}

\graphicspath{{figs/},{paper\_figs/}}

\usepackage{hyperref}
\hypersetup{ 
	colorlinks   =  true
}

\begin{document}
	
\title{Broken symmetry in a two-qubit quantum control landscape}%

\author{Marin Bukov}
\email{mgbukov@berkeley.edu}
\affiliation{Department of Physics, University of California, Berkeley, CA 94720, USA}
\affiliation{Department of Physics, Boston University, 590 Commonwealth Ave., Boston, MA 02215, USA}

\author{Alexandre G.~R. Day}
\email{agrday@bu.edu}
\affiliation{Department of Physics, Boston University, 590 Commonwealth Ave., Boston, MA 02215, USA}

\author{Phillip Weinberg}
\affiliation{Department of Physics, Boston University, 590 Commonwealth Ave., Boston, MA 02215, USA}

\author{Anatoli Polkovnikov}
\affiliation{Department of Physics, Boston University, 590 Commonwealth Ave., Boston, MA 02215, USA}

\author{Pankaj Mehta}
\affiliation{Department of Physics, Boston University, 590 Commonwealth Ave., Boston, MA 02215, USA}

\author{Dries Sels}
\affiliation{Department of Physics, Boston University, 590 Commonwealth Ave., Boston, MA 02215, USA}
\affiliation{Department of Physics, Harvard University, 17 Oxford st., Cambridge, MA 02138, USA}
\affiliation{Theory of quantum and complex systems, Universiteit Antwerpen, B-2610 Antwerpen, Belgium}

\begin{abstract}
	We analyze the physics of optimal protocols to prepare a target state with high fidelity in a symmetrically coupled two-qubit system. By varying the protocol duration, we find a discontinuous phase transition, which is characterized by a spontaneous breaking of a $\mathbb{Z}_2$ symmetry in the functional form of the optimal protocol, and occurs below the quantum speed limit. We study in detail this phase and demonstrate that even though high-fidelity protocols come degenerate with respect to their fidelity, they lead to final states of different entanglement entropy shared between the qubits. Consequently, while globally both optimal protocols are equally far away from the target state, one is locally closer than the other. An approximate variational mean-field theory which captures the physics of the different phases is developed. 
\end{abstract}

\date{\today}
	
	
\maketitle

\section{\label{sec:intro}Introduction}

The most rudimentary characterization of matter is arguably in terms of its thermodynamic phase, such as liquid, solid and gas, with each phase featuring its own distinct macroscopic properties.  Whether a system is in one phase or the other is determined by a combination of intrinsic microscopic parameters (coupling constants) and some macroscopic parameters, such as temperature or pressure.  

\begin{figure}[t!]
	\includegraphics[width=1.0\columnwidth]{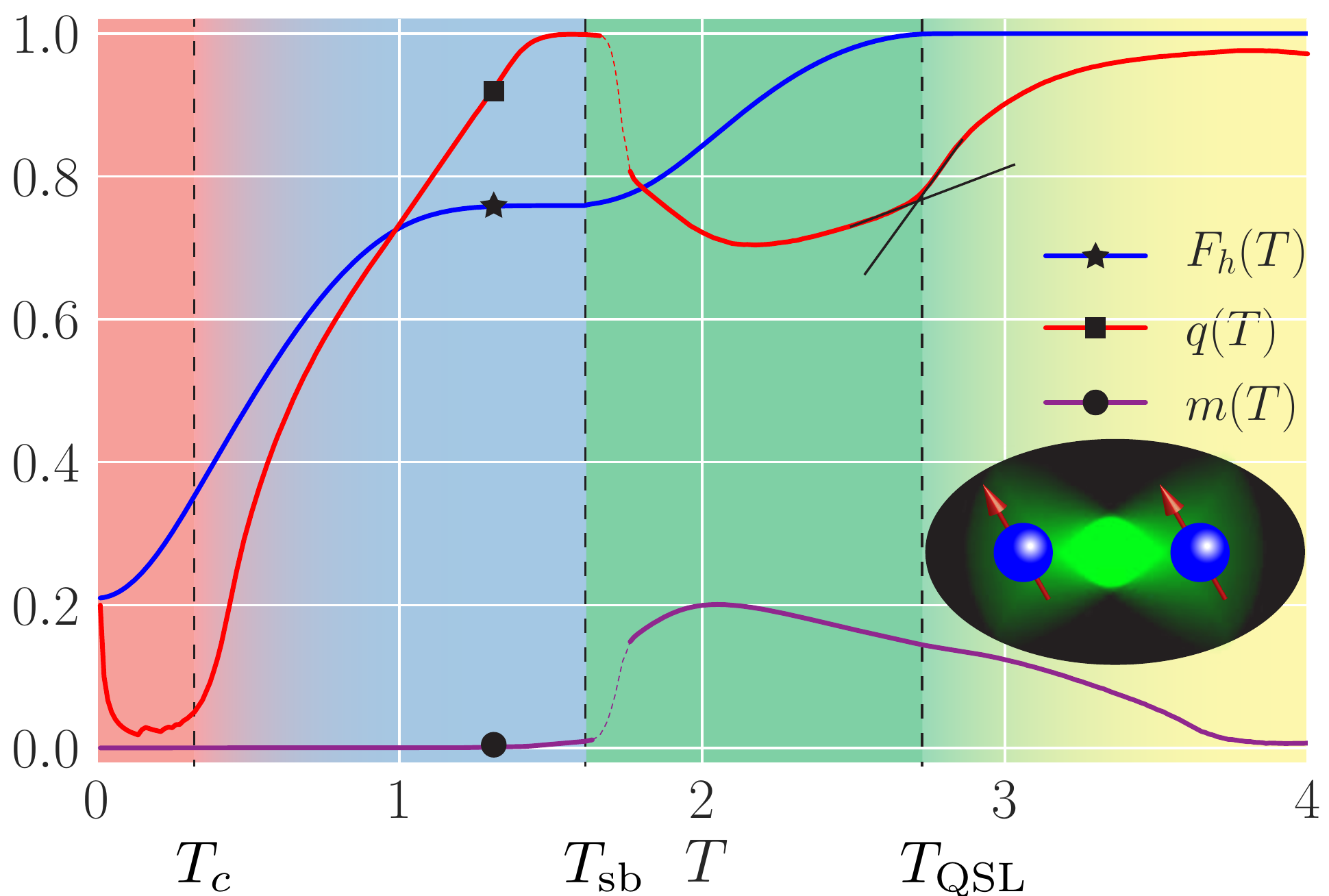}
	\caption{\label{fig:phase_diag_2B}Quantum control phase diagram for the symmetrically-coupled two-qubit problem (inset). As a function of the protocol duration $T$ the infidelity landscape exhibits both continuous and discontinuous phase transitions (vertical dashed lines) featuring overconstrained (red), correlated/glassy (blue and green), symmetry-broken (green) and controllable (yellow) phases, detected by the correlator $q(T)$. The fidelity $F_h(T)$ of the optimal protocol and the corresponding magnetization order parameter $m(T)$, feature distinctive behaviour in the different phases with non-analyticities at the phase boundaries emerging in the limit of vanishing protocol time step size. Time is in units of the inverse qubit interaction strength.}
\end{figure}

In direct analogy, we find that the process of preparing states in quantum systems can be characterized in different phases, each phase having a distinct feature, c.f.~Fig.~\ref{fig:phase_diag_2B}. Whether the control problem belongs to a certain phase depends on the details of the underlying quantum system, as well as on a global external parameter -- the protocol duration. Consequently, by varying the protocol duration, the control problem can change the phase. In much the same way conventional phase transitions carry far-reaching consequences for understanding the properties of physical substances, the quantum control phase transitions play a quintessential role for manipulating quantum states with high efficiency.

In this paper, we report on a discrete symmetry breaking in the state preparation problem of a two-qubit system. A key role for the existence of this phase seems to be played by quantum entanglement. This helps us construct an effective approximate variational theory, which captures the essential features of the optimal protocol, and the physics of the control phase transitions. 

The difficulty underlying quantum state preparation is inherited from its intrinsically non-equilibrium character, and the question of whether efficient state preparation is feasible in many-body systems remains largely open. The ability to prepare target states quickly and with high fidelity is central to the study and manipulation of quantum mechanical systems, and constitutes a major bottleneck in various cutting-edge modern studies: quantum computing~\cite{nielsen} relies vastly on the capability of transferring the population with high fidelity from an initial to a target state; experiments with ultracold atoms~\cite{bason_12,vanfrank_16,wigley_16}, trapped ions~\cite{islam_11,senko_15,jurcevic_14}, superconducting qubits~\cite{barends_16}, and NV centres~\cite{zhou_17}, have to first prepare the system in the desired state, in order to explore the interesting physics hidden in it.

Adiabatic processes in complex many-body systems may require very long preparation times which are often not affordable in practice. This inspired the development of the theory of counter-diabatic and fast-forward driving~\cite{vitanov_96,demirplak_05,berry_09,masuda_09,jarzynski_13,torrontegui_13,delcampo_13,deffner_14,kolodrubetz_16,sels_16,baksic_16,patra_17,jarzynski_17,bukov_GSL}, which exploits the nonequilibrium features of the problem to design protocols leading to transitionless driving.
At the same time, in the era of computation, optimal control theory has been developed to address state preparation as an optimisation problem~\cite{rabitz_98,glaser_98,lloyd_14,glaser_15,zhdanov_17}. Prominent algorithms, such as gradient-based CRAB~\cite{caneva_11} and GRAPE~\cite{grape_05}, and model-free Machine Learning~\cite{judson_92,chen_14,chen_14_ML,bukov_17RL,yang_17,dunjko_17,ML_review} have recently been successfully applied to find (nearly) optimal protocols in quantum many-body systems. 

This paper is organized as follows. In Sec.~\ref{sec:model} we define the two-qubit control problem, analyze its control phase diagram with emphasis on symmetry breaking in Sec.~\ref{sec:phase_diag}, and introduce a `magnetization' order parameter in Sec.~\ref{sec:mag_OP}. In Sec.~\ref{sec:correlations} we study extensively the correlations between the local minima of the control landscape (i.e.~the nearly-optimal fidelity protocols) across the control critical points. We proceed with Sec.~\ref{sec:ent_obs}, where we discuss the consequences of symmetry breaking for physical observables and the entanglement entropy shared between the qubits. In Sec.~\ref{sec:variational} we develop an effective variational theory for the control phase diagram. Finally, we conclude in Sec.~\ref{sec:outro}. 

\section{\label{sec:model}Model}

We study the physics of optimal protocol sequences which attempt at preparing a target state in a symmetrically-coupled two-qubit system. This model represents the simplest non-trivial generalisation of the exactly solvable two-level system. The state preparation problem in this deceptively simple system lacks a closed-form analytical solution, and exhibits a remarkably rich control phase diagram, c.f.~Fig.~\ref{fig:phase_diag_2B}. The Hamiltonian is
\begin{equation}
\label{eq:H_2B}
H[h_x(t)] = -2JS^z_1S^z_2 - g (S_1^z + S_2^z) - h_x(t)(S_1^x + S_2^x),
\end{equation}
where $J\!=\!g\!=\!1$ are the interaction strength and the static magnetic field along the $z$-axis, respectively, and $h_x(t)$ is the time-dependent control field. The Pauli spin-$1/2$ operators are denoted by $S^\mu_{j=1,2}$. We prepare the system at time $t=0$ in the ground state (GS) $|\psi_i\rangle$ of $H[h_x=-2]$ and want to transfer the population into the target state $|\psi_*\rangle$ -- the GS of $H[h_{x,\ast}=+2]$ -- in a fixed amount of time $T$. Thus, our goal is to find the functional form of the driving protocol $h_x(t)$ ($t\in[0,T]$) which maximizes the fidelity of being in the target state $F_h(T)=|\langle\psi(T)|\psi_*\rangle|^2$. Here $|\psi(T)\rangle$ denotes the final state at $t=T$, following a unitary Schr\"odinger evolution for a duration $T$. 

Notice that the Hamiltonian is invariant under exchanging the two qubits and, therefore, the Hilbert space factorizes into a triplet and a singlet manifold. The GS $\vert\psi_i\rangle$ belongs to the triplet manifold, to which the dynamics is confined, since the control field respects this qubit-exchange symmetry at all times. Hence, the above problem effectively reduces to a three-level system, with the space of all possible operators spanned by $SU(3)$. Optimal transfer of population from the GS to the highest-energy state (a.k.a.~pumping) in driven three-level systems has been studied using Lie group methods~\cite{jurdjevic_72,boscain_02,boscain_04} and a closed-form solution has been derived. Three-level systems have also been studied using ideas form shortcuts to adiabaticity~\cite{chang_07,sugny_08,chen_12,giannelli_14,li_16,theisen_17}.

Observe that the state preparation optimisation problem outlined above has a hidden discrete symmetry. Since
\begin{equation}
\mathrm e^{i\pi (S^z_1+S^z_2)}H[J,g,h_x]\mathrm e^{-i\pi (S^z_1+S^z_2)}=H[J,g,-h_x],
\end{equation}
it follows that 
\begin{equation}
\mathrm e^{-i\pi (S^z_1+S^z_2)}|\psi_i\rangle\!=\!|\psi_\ast\rangle.
\end{equation}
Denoting by $U_{h(t)}(T,0)$ the evolution operator between times $0$ and $T$ following the protocol $h_x(t)$, it is straightforward to show $F_{h(t)}(T) = F_{-h(T-t)}(T)$:
\begin{eqnarray}
F_{h(t)}(T)&=&\vert\langle\psi_\ast\vert U_{h(t)}(T,0)\vert\psi_i\rangle\vert^2 \nonumber\\
&=&\vert\langle\psi_i\vert \mathrm{e}^{+i\pi (S^z_1+S^z_2)} U_{h(t)}(T,0) \mathrm{e}^{-i\pi (S^z_1+S^z_2)} \vert\psi_\ast\rangle\vert^2 \nonumber\\
&=&\vert\langle\psi_i\vert U_{-h(t)}(T,0) \vert\psi_\ast\rangle\vert^2 \nonumber\\
&=&\vert\langle\psi_\ast\vert \left[U_{-h(t)}(T,0)\right]^\dagger \vert\psi_i\rangle\vert^2  \\
&=&\vert\langle\psi_\ast \vert U_{-h(t)}(0,T) \vert\psi_i\rangle\vert^2 \nonumber\\
&=&\vert\langle\psi_\ast\vert U_{-h(T-t)}(T,0)\vert\psi_i\rangle\vert^2 = F_{-h(T-t)}(T), \nonumber
\label{eq:symmetry}
\end{eqnarray}
for any protocol $h_x(t)$. Hence, the optimal protocol is either unique, obeying the discrete $\mathbb{Z}_2$ symmetry $h_x(t)\!=\!-h_x(T\!-\!t)$ or, since the symmetry group is $\mathbb{Z}_2$, it is doubly degenerate and breaks this symmetry.

\begin{figure*}[t!]
	\includegraphics[width=1.0\textwidth]{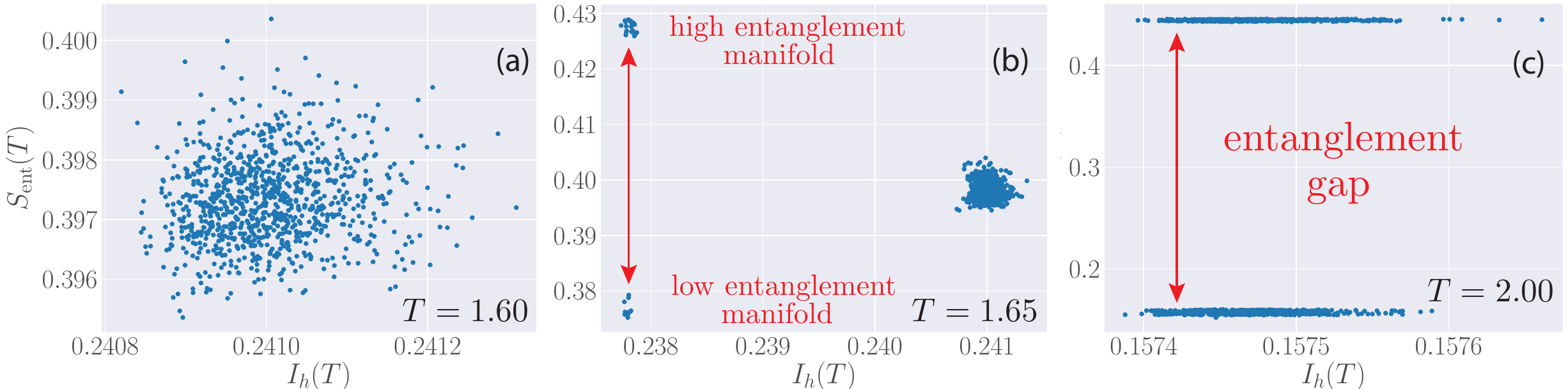}
	\caption{\label{fig:Send_vs_Fid} Entanglement entropy $S_\mathrm{ent}(T)$ against infidelity $I_h(T)$ for the final state at three different protocol durations $T$ across the symmetry-breaking transition. Each data point corresponds to evolution following a protocol $h_\alpha(t)$ from the set of local infidelity minima obtained using SD. An entanglement gap between two large clusters is present in the symmetry-broken phase. The protocol step size is fixed at $\delta t = 0.00125$. Time is in units of the inverse qubit interaction strength.}
\end{figure*}

\section{\label{sec:phase_diag}Quantum Control Phase Diagram}

The quantum speed limit (QSL) in the context of optimal control is defined as the minimal time $T_\mathrm{QSL}$ required to prepare the target state with strictly \emph{unit} fidelity. In generic problems, where one only has a limited control over the degrees of freedom, and where the control fields strength are bounded, $T_\mathrm{QSL}>0$. For gapless many-body systems, it is expected that $T_\mathrm{QSL}\to\infty$ in the thermodynamic limit. The existence of a finite QSL renders a system controllable~\cite{jurdjevic_72}.

For the Hamiltonian~\eqref{eq:H_2B}, the existence of a finite quantum speed limit follows from general theorems about control systems on compact Lie groups~\cite{jurdjevic_72}, and the fact that repeated nested commutators of the non-driven $H_0=-2S^z_1S^z_2 -(S_1^z + S_2^z)$ and driven $H_1=-(S_1^x + S_2^x)$ parts of the Hamiltonian, generated during the time evolution, exhaust the entire operator manifold $SU(3)$. Unfortunately, the proofs of these existence theorems are non-constructive, and hence they appear to be of limited use in experimental and numerical studies. Nevertheless, we were able to identify a simple variational symmetric three-pulse sequence, which yields unit fidelity in a finite time, see App.~\ref{app:var_control}. The minimal protocol duration for reaching unit fidelity \emph{within this variational family} of protocols immediately puts an upper bound on $T_\mathrm{QSL}$ of approximately $T_\mathrm{QSL} < 2.907$.

For $T\!<\!T_\mathrm{QSL}$, there exists no protocol to prepare the target state $\vert\psi_\ast\rangle$ with unit fidelity. Nevertheless, the question of what the optimal protocol and the corresponding fidelity are, is of particular interest since, for generic many-body problems, $T_\mathrm{QSL}$ is typically very large (if at all finite), and one is virtually always forced to work in this regime. 

Assuming we have limited resources available, we study the highly-constrained problem~\eqref{eq:H_2B} of a \emph{single} global control $x$-field of \emph{bounded} strength $\vert h_x(t)\vert \leq 4$. Pontryagin's maximum principle implies that there exists an optimal protocol which only takes values on the boundary of the allowed domain for almost all times. Thus, we choose to restrict to bang-bang protocols, defined by $h_x(t)\in\{\pm 4\}$, with a total of $N_T$ steps of size $\delta t$~\cite{OC_book}. We verified that our conclusions remain unchanged if we consider continuous protocols. It has recently been shown that this control problem is equivalent to finding the lowest-energy configuration of a highly nonlocal, frustrated classical Ising spin model with energy $\mathcal{H}_\mathrm{eff}(T)$, which features all-to-all multi-body interactions~\cite{day_17}. Note that the involved classical spin degrees of freedom correspond to the bangs in the protocol $h_x(t)$, and are distinct from the quantum spins $S_j^\mu$. Even though the original system may have only a few \emph{quantum} degrees of freedom, $\mathcal{H}_\mathrm{eff}(T)$ describes a complex interacting many-body system~\cite{day_17}. To better appreciate this analogy, notice that any bang-bang protocol $h_x(t)$ can be uniquely mapped to a classical Ising spin configuration. To each such classical spin state $h_x(t)$, we can assign as `energy' its infidelity value $h_x(t)\mapsto I_h(T)=1-F_h(T)$. Determining the optimal protocol then corresponds to finding the minimum of the infidelity landscape, i.e.~the lowest-infidelity spin configuration of $\mathcal{H}_\mathrm{eff}(T)$. 

\begin{figure*}[t!]
	\includegraphics[width=1.0\textwidth]{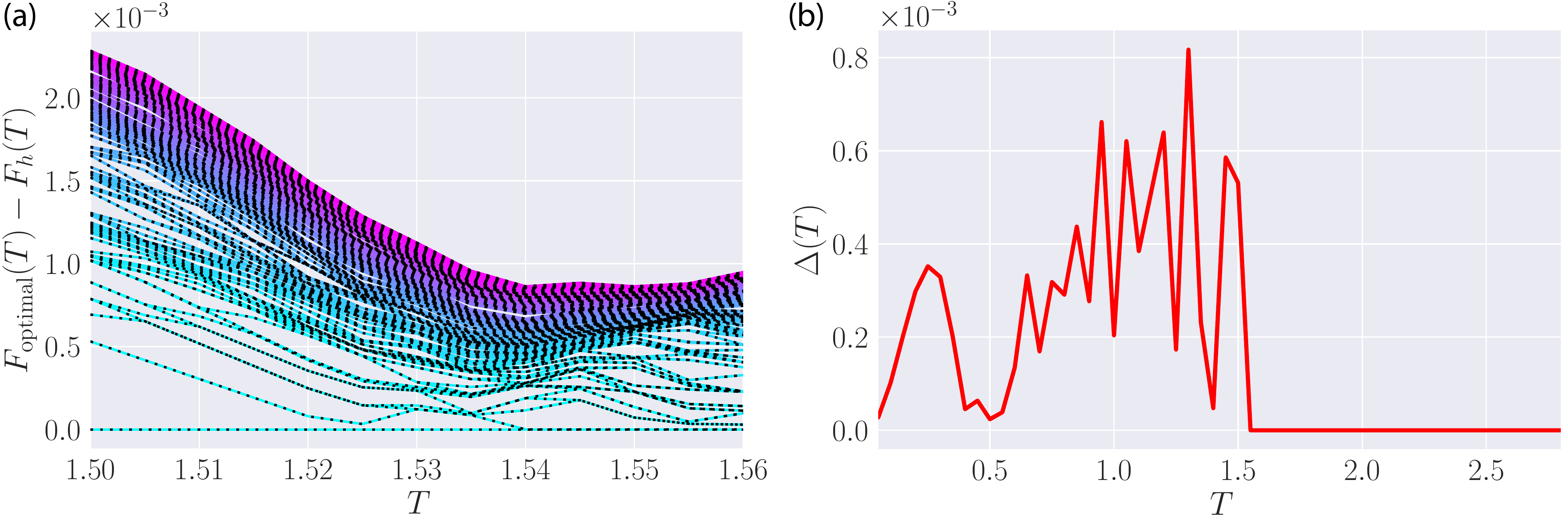}
	\caption{\label{fig:spectrum} (a) Low-infidelity manifold of the optimisation landscape as a function of the protocol duration (the optimal fidelity is shifted to zero for every $T$). The colours indicate the infidelity density. (b) The infidelity gap $\Delta$ as a function of $T$ vanishes at the symmetry-breaking point $T_\mathrm{sb}$ and the optimal protocol becomes degenerate. The number of bangs in a protocol is kept fixed at $N_T=28$. Time is in units of the inverse qubit interaction strength.}
\end{figure*}

There exists a one-to-one correspondence between the thermodynamic limit for this classical model and the limit of vanishing protocol step size: $\delta t\to 0$, $N_T\to\infty$ with $T=\delta tN_t=const$. Interestingly, the classical many-body system described by $\mathcal{H}_\mathrm{eff}(T)$ features a variety of low energy phases as a function of the protocol duration $T$. In order to reveal them, we use Stochastic Descent (SD)~\cite{bukov_17RL} to obtain a set of $N_\mathrm{real}$ local infidelity minima $\{h_x^\alpha(t)\}_{\alpha=1}^{N_\mathrm{real}}$. If we denote by $\overline{h_x}(t)=N_\mathrm{real}^{-1}\sum_{\alpha=1}^{N_\mathrm{real}}h_x^\alpha(t)$ the statistical average over this set at a fixed time $t$, we can define a correlator between the protocols as
\begin{equation}
q(T)=\frac{1}{16N_T}\sum_{j=1}^{N_T}\overline{\{h_x(j\delta t)-\overline{h_x}(j\delta t)\}^2},
\label{eq:q_EA}
\end{equation}
which is closely related to the Edwards-Anderson order parameter used to measure spin-glass order~\cite{castellani_05,hedges_09,bukov_17RL}. Whenever the local infidelity minima $\{h^\alpha(t)\}_{\alpha=1}^{N_\mathrm{real}}$ are completely uncorrelated, we have $q(T)\equiv 1$, while for a convex infidelity landscape -- $q(T)\equiv 0$.

Figure~\ref{fig:phase_diag_2B} shows the phase diagram of this quantum control problem, as determined by the correlation function $q(T)$. Starting at protocol times $T\approx 3$, we find the optimal fidelity (blue line) at unity, which means that one can successfully and completely prepare the target state $|\psi_\ast\rangle$.  Therefore, the system is said to be in the \emph{controllable} phase (yellow). 

At the critical point $T_\mathrm{QSL}\approx 2.8$, the infidelity landscape undergoes a continuous phase transition to a correlated glassy phase (blue, green). One can think of this critical point as a phase transition in the effective classical spin model $\mathcal{H}_\mathrm{eff}$. For $T<T_\mathrm{QSL}$, the fidelity $F_h(T)$ deviates from unity, and reaching the target state becomes impossible under the constraints of the problem. We emphasize that this is a sharp transition from strictly unit fidelity, and not just a crossover behaviour [see finite size scaling in App.~\ref{app:scaling}]. In this glassy phase, the protocols associated with local minima of the infidelity landscape become correlated, which is reflected in a value of the order parameter $q(T)$ less than unity. Due to the glassiness in the infidelity landscape, sophisticated algorithms with nonlocal updates are required to look for the optimal protocol [a.k.a.~the global minimum].

\begin{figure*}[t!]
	\includegraphics[width=1.0\textwidth]{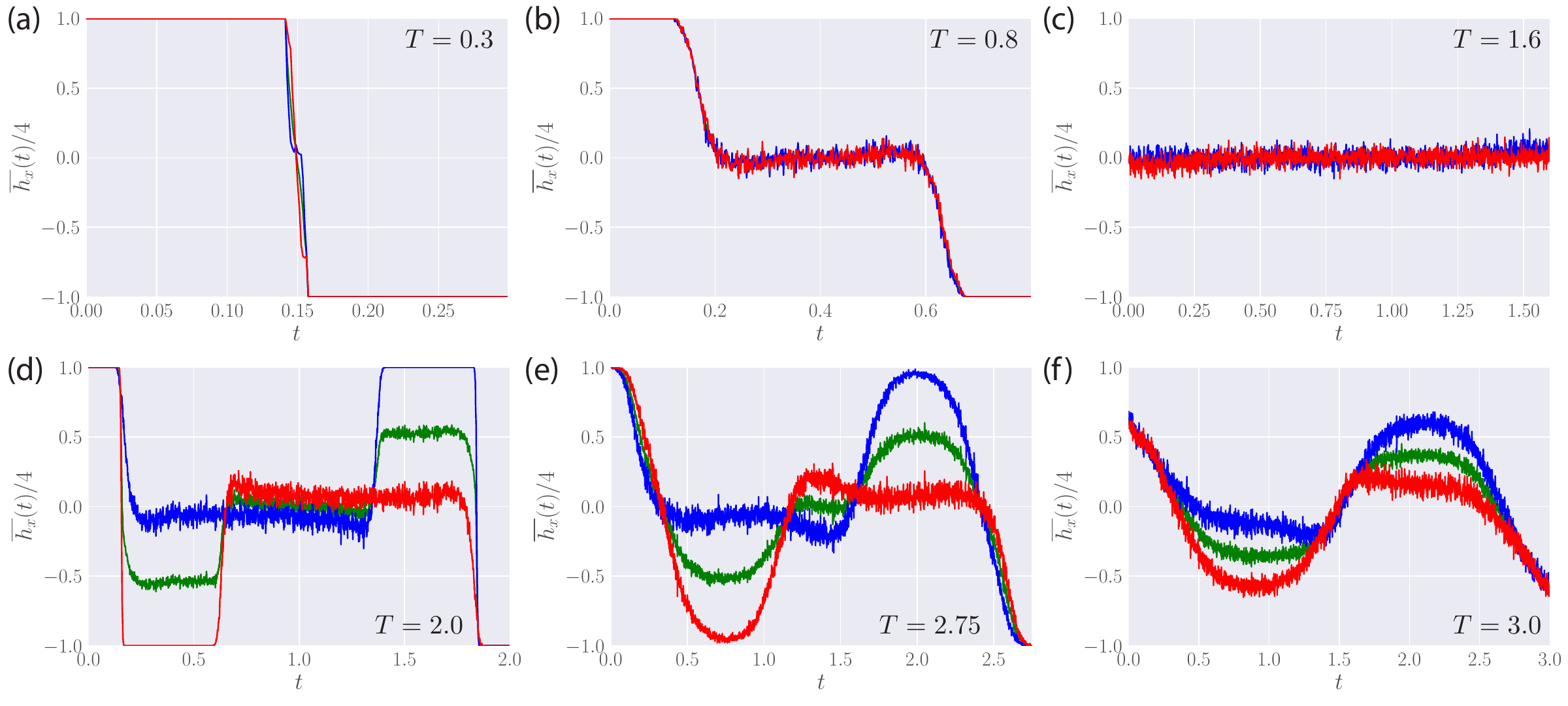}
	\caption{\label{fig:protocol_averaged} The set of $10^3$ protocols (sample of local infidelity minima obtained using SD) can be divided into two disjoint subsets according to the mean entanglement entropy (see Fig.~\ref{fig:Send_vs_Fid}), leaving a total of three sets: the low-$S_\mathrm{ent}$ protocols (red, down), the high-$S_\mathrm{ent}$ protocols (blue, up) and all protocols (green, middle). Symmetry breaking in the low-infidelity manifold of the control landscape becomes evident. The time step size $\delta t=0.00125$. Time is in units of the inverse qubit interaction strength.}
\end{figure*}

The correlated phase of the system~\eqref{eq:H_2B} itself consists of two other phases: (i) for $T\gtrsim T_\mathrm{sb}\approx 1.57$, spontaneous symmetry breaking occurs in protocol space. In the language of the effective many-body classical spin model $\mathcal{H}_\mathrm{eff}$, the broken discrete $\mathbb{Z}_2$ symmetry, c.f.~Eq.~\eqref{eq:symmetry}, is equivalent to reflection about the centre of the time lattice followed by a global classical-spin inversion. Symmetry breaking is also observed in the exact infidelity landscape of a system of $N_T=28$ bangs, see Fig.~\ref{fig:spectrum}. At the critical point, the low-infidelity manifold splits in two distinct sets of protocols. These sets contain protocols equivalent w.r.t.~their fidelity, but separated by a finite gap in the entanglement entropy $S_\mathrm{ent}(T)$ they create in the evolved state, see Fig.~\ref{fig:Send_vs_Fid}. Precisely at the symmetry breaking critical point $T_\mathrm{sb}$, the entanglement gap between the two sets closes, lifting the distinction between protocols, and the low-infidelity manifold of the control landscape becomes completely uncorrelated and symmetric. This behaviour is accompanied by a jump in $q(T)$ and the magnetisation order parameter $m(T)$ in the limit $\delta t\to 0$ [see finite size scaling in App.~\ref{app:scaling}], and hence the transition is discontinuous, at least within the family of bang-bang protocols. The optimal protocol is symmetric for $T<T_\mathrm{sb}$, and symmetry-broken for $T>T_\mathrm{sb}$. Despite the transition being discontinuous, we find that at the critical point the optimal protocol is $h_x\equiv 0$, which is both symmetric and antisymmetric. This means that, at the symmetry-breaking point, the optimal strategy is to completely turn off the driving field and simply wait. Using this fact, we were able to determine that, for the optimal protocol, $T^{h_\mathrm{optimal}}_\mathrm{sb}=\pi/2$. The simplicity of this expression is a consequence of setting $J=g$, cf.~App.~\ref{app:Tsb_derivation}. After averaging over the sample $\{h_x^\alpha(t)\}_{\alpha=1}^{N_\mathrm{real}}$, the true value for the transition, as detected by the order parameter $q(T)$, is most likely somewhere in the vicinity, i.e.~$T_\mathrm{sb}\approx\pi/2$. Because the sample-average protocol $\overline{h}_x(t)\equiv 0$ for $T=T_\mathrm{sb}$ is both even and odd [see Fig.~\ref{fig:protocol_averaged}c], it allows to smoothly change symmetry, indicating that the transition might become continuous if we do not restrict the protocols to the bang-bang family.
Approaching the critical point from below, (ii), we have $q(T\!\to\! T_\mathrm{sb}^-)=1$, and hence the protocols at $T_\mathrm{sb}$ are completely uncorrelated. 

Lowering the total protocol duration $T$ further, we encounter yet another continuous phase transition at $T\!=\!T_c\!\approx\! 0.38$, when the various minima of the infidelity landscape coalesce into a single global minimum, and $q(T)\!=\!0$ in the limit $\delta t\to 0$. This suggests that the landscape in this \emph{overconstrained} phase (red) is convex, and optimisation is easy again, even though the optimal fidelities one can reach are relatively poor due to the short protocol duration. 

A similar, dynamical symmetry-breaking phenomenon was reported in Ref.~\cite{kappen2005linear} in the case of stochastic optimal control. While there is a number of similarities in the two concepts, the phenomenon of Ref.~\cite{kappen2005linear} relies exclusively on the stochastic nature of the problem considered therein, while our setup is completely deterministic. Moreover, symmetry breaking in Ref.~\cite{kappen2005linear} is dynamical and appears in physical time, while in our case it happens as we vary the total protocol duration $T$.

\section{\label{sec:mag_OP}Magnetisation Order Parameter for Symmetry Breaking in the Control Landscape}

The symmetry-broken correlated phase can also be detected by a suitable order parameter, which we now construct. The `magnetization' of the protocol $h_x(j\delta t)$, viewed as a classical spin state, is 
\begin{equation}
m_h(T) = \frac{1}{4N_T}\sum_{j=1}^{N_T}h_x(j\delta t). 
\end{equation}
Similar to the other control phase transitions, this discontinuous transition occurs at finite infidelity density [energy density of $\mathcal{H}_\mathrm{eff}$], since its appearance can be seen in the entire low-infidelity part of the glassy spectrum, not just the optimal protocol, cf.~Fig.~\ref{fig:Send_vs_Fid}. To reveal this, we define the minima-averaged magnetisation
\begin{equation}
\label{eq:magnetisation_OP}
m(T) = \frac{1}{N_\mathrm{real}} \sum_{\alpha=1}^{N_\mathrm{real}} |m_{h^{\alpha}}(T)|.
\end{equation}
Figure.~\ref{fig:phase_diag_2B} shows that both the correlator $q(T)$, and the magnetisation order parameter $m(T)$ feature jumps precisely at $T\!=\!T_\mathrm{sb}$, which sharpen with decreasing the time step size [see finite size scaling in App.~\ref{app:scaling}]. We can, therefore, deduce that the symmetry-breaking transition is discontinuous, at least within the family of bang-bang protocols.

Since it is impossible to reliably obtain the exact low-infidelity part of the control landscape we resort to an exhaustive search, in order to study the symmetry-breaking phenomenon more closely. We fix a total of $N_T=28$ bangs and compute all $2^{28}$ protocols and their fidelities. Figure~\ref{fig:spectrum}a shows the best fidelities in the region of the symmetry-breaking phase transition. One can clearly see how the GS and the first excited state merge into a degenerate doublet, while the excitations follow a similar behaviour. Another manifestation of this is displayed in Fig.~\ref{fig:spectrum}b which shows that the gap between the best and second-best protocols (a.k.a.~the GS and the first excited state of $\mathcal{H}_\mathrm{eff}$), vanishes completely for $T_\mathrm{sb}<T$. It is an interesting observation that different states do not undergo symmetry breaking simultaneously, although it is an open question whether this is due to the finite size of the protocol time step. Nevertheless, one can clearly identify the level crossings leading to a drastic reorganisation of the involved protocols w.r.t.~their infidelity close to $T_\mathrm{sb}$.

\section{\label{sec:correlations}Correlations between Local Minima of the Control Landscape}

The order parameter for detecting a quantum control phase transitions, $q(T)$, measures the correlations between local minima of the infidelity landscape averaged over time. In this section, we resolve the time-dependence of these correlations and study their behaviour as a function of the protocol duration $T$.

Let us define the connected protocol-protocol correlator as
\begin{eqnarray}
C(t,t') &=& \frac{1}{16}\overline{\{h_x(t)-\overline{h_x}(t)\}\{h_x(t')-\overline{h_x}(t')\}} \\
&=& \frac{1}{16N_\mathrm{real}}\sum_{\alpha=1}^{N_\mathrm{real}}h_x^\alpha(t)h_x^\alpha(t') - \overline{h_x}(t)\overline{h_x}(t'), \nonumber
\label{eq:correlators}
\end{eqnarray}
where the averaging $\overline{h_x}(t)$, as before, is done over the set of local infidelity minima $\{h_x^\alpha\}_\alpha$ at fixed time, and the factor $1/16$ serves to normalise each protocol to $\{\pm 1\}$. This quantity measures the fluctuations about the mean of nearly-optimal fidelity protocols and is, therefore, sensitive to phase transitions where drastic changes in the infidelity landscape occur. We distinguish between equal-time and non-equal time correlations. A value of $C(t,t')=1$ suggests a complete absence of correlation amid almost optimal protocols. 

\begin{figure*}[t!]
	\includegraphics[width=1.0\textwidth]{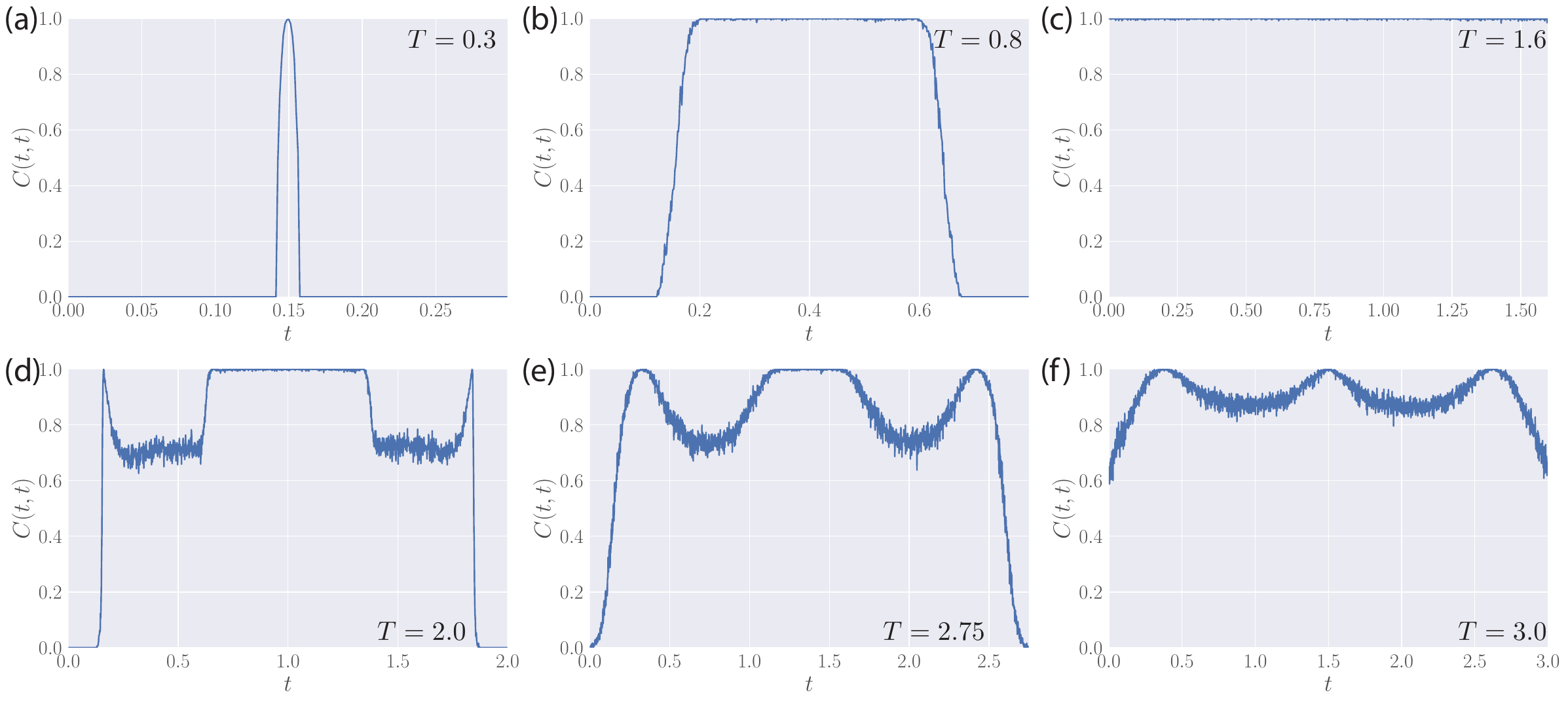}
	\caption{\label{fig:equatime_corr}Equal-time correlations in the low-infidelity landscape as a function of time $t$. The averaging is done over a sample of $10^3$ local infidelity minima. The time step size $\delta t=0.00125$. Time is in units of the inverse qubit interaction strength.}
\end{figure*}

While the Edwards-Anderson-like order parameter can be obtained from the time averaged equal-time correlator:
\begin{equation}
q(T) = \frac{1}{N_T}\sum_{j=1}^{N_T}C(j\delta t,j\delta t),
\end{equation}
non-equal time correlations contain further information about the structure of the control landscape, which can be understood intuitively as follows. Since we are studying a dynamical problem, the correlations in the protocols arise primarily due to two reasons: (i) causality which is imposed by Schr\"odinger evolution suggests that the value of $h_x(t')$ at time $t'$ depends on the values of the protocol at all previous times $t<t'$. (ii) The underlying symmetry of the control problem imposes further correlations between the points $t<T/2$ and $t>T/2$.

\begin{figure*}[t!]
	\includegraphics[width=1.0\textwidth]{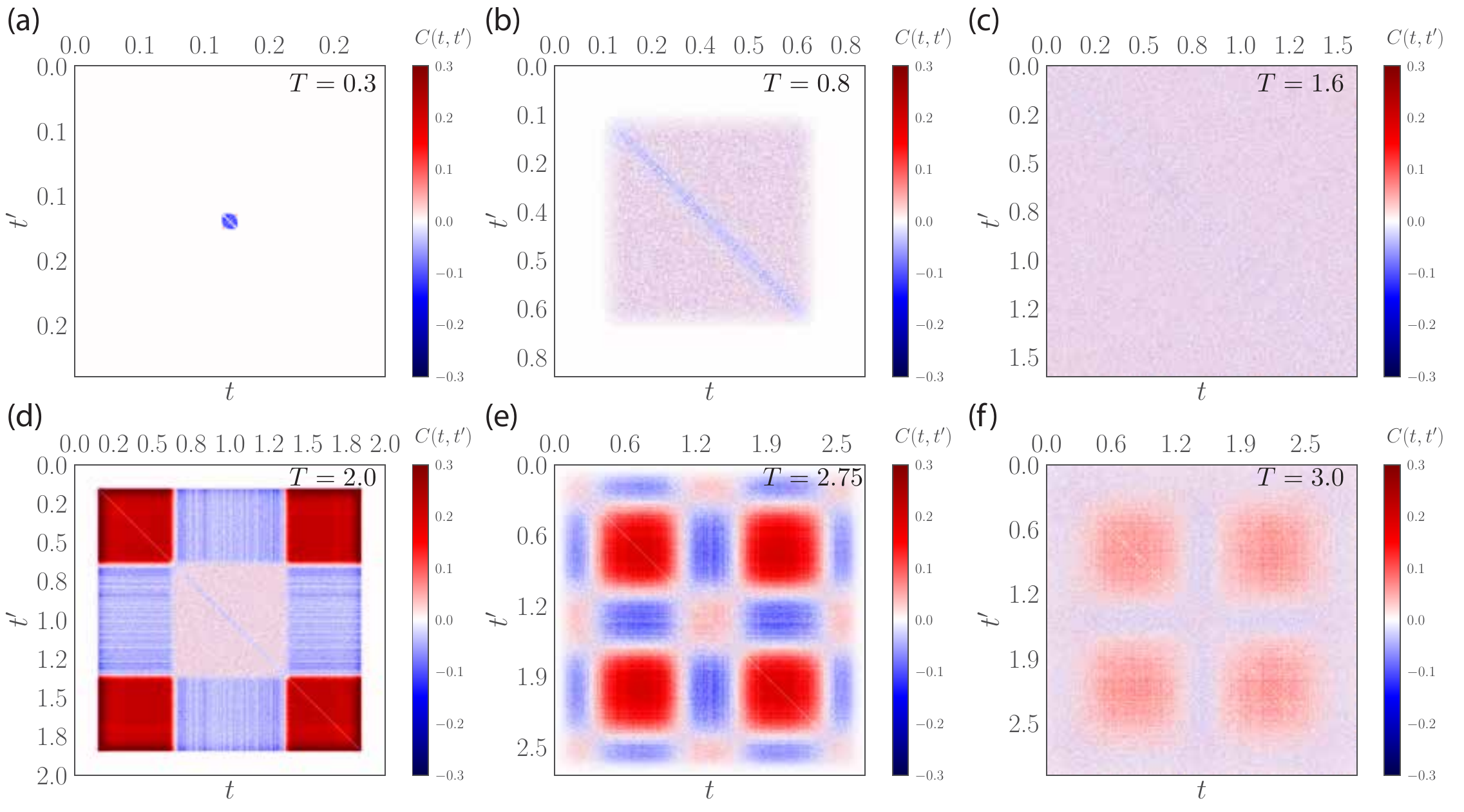}
	\caption{\label{fig:nonequatime_corr}Non-equal time correlations in the low-infidelity landscape as a function of time $t$. The averaging is done over a sample of $10^3$ local infidelity minima. The time step size $\delta t=0.00125$. Time is in units of the inverse qubit interaction strength.}
\end{figure*}

Notice that we explicitly subtracted the mean values $\overline{h}(t)$ from the definition in Eq.~\eqref{eq:correlators}. The sample-averaged protocol $\overline{h}(t)$ reveals information about the structure of local attractors in the infidelity landscape. Figures~\ref{fig:protocol_averaged},~\ref{fig:equatime_corr} and~\ref{fig:nonequatime_corr} show the sample-averaged protocols $\overline{h}(t)$, and the equal and non-equal time correlators for different protocol durations $T$. Observe how the effective number of degrees of freedom (i.e.~number of independent pulse lengths) in the protocol changes from one at $T<T_c$, to two for $T_c<T<T_\mathrm{sb}$ in the symmetric correlated phase. As anticipated, symmetry breaking becomes manifest in the symmetry-broken glass phase for $T_\mathrm{sb}<T<T_\mathrm{QSL}$ where the averaged protocol has three independent degrees of freedom. We find that both correlators are also sensitive to the discontinuous symmetry-breaking transition, and feature sharp changes at $T\approx T_\mathrm{sb}$.

\section{\label{sec:ent_obs}Entanglement and Observables in the Symmetry-Broken Phase}

\begin{figure*}[t!]
	\includegraphics[width=1.0\textwidth]{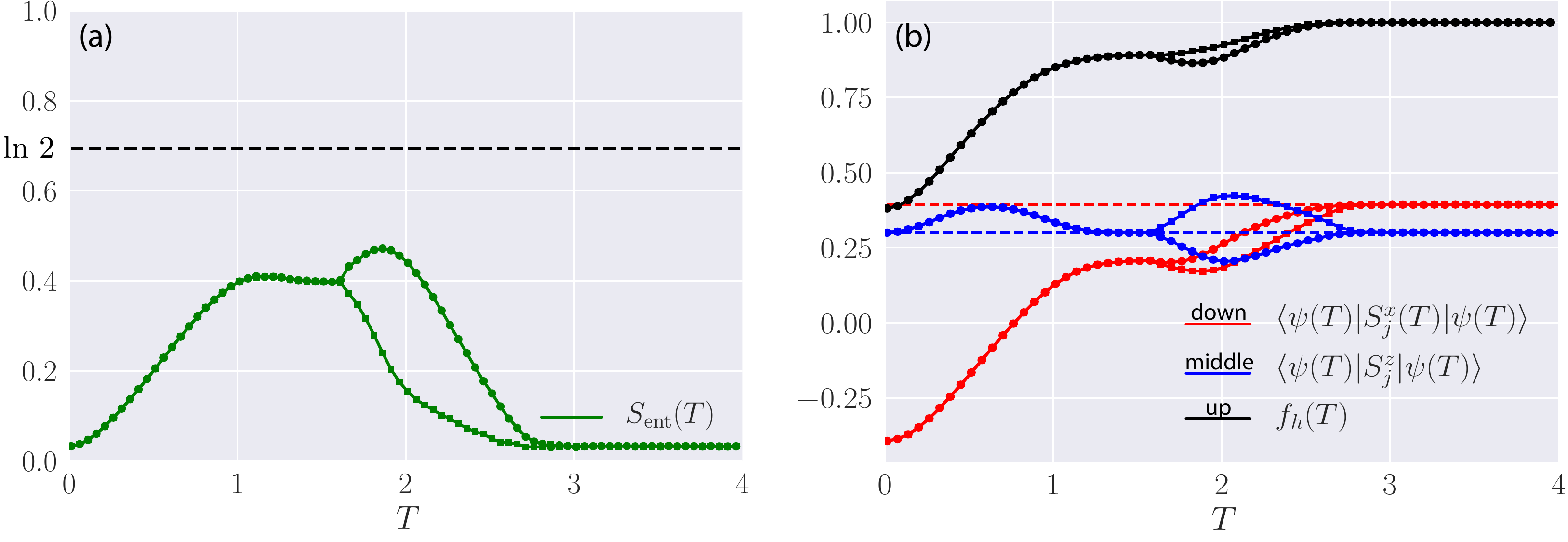}
	\caption{\label{fig:Sent_split} (a) Entanglement entropy $S_\mathrm{ent}(T)$ shared between the qubits and (b) local observables after evolution with the best protocol(s) found using SD as a function of the protocol duration $T$ for $\delta t = 0.00125$. A bifurcation is visible in the symmetry-broken phase where the optimal protocol is doubly degenerate. Time is in units of the inverse qubit interaction strength.}
\end{figure*}

In the correlated (glassy) phase, the best fidelity is no longer unity, and the optimal protocol leads to a final state, which is different from the target state. An intriguing question to ask is how much entanglement the optimal protocol creates. Tracing out one of the two qubits, we can measure the shared entanglement entropy
\begin{equation*}
S_\mathrm{ent}(T) = -\mathrm{tr}\left(\rho_1\ln\rho_1\right),\qquad \rho_1=\mathrm{tr}_2\vert\psi(T)\rangle\langle\psi(T)\vert
\end{equation*} 
at the end of the protocol at time $T$, shown in Fig.~\ref{fig:Sent_split}(a) (green line) and Fig.~\ref{fig:Send_vs_Fid} for the entire low-infidelity sample. Notice how the degeneracy in the low-infidelity manifold shows up as a bifurcation in the entanglement entropy curve throughout the entire symmetry-broken glass phase (Fig.~\ref{fig:phase_diag_2B}, green). This phenomenon occurs because the entanglement entropy is not invariant under the symmetry of the protocol $h_x(t)\mapsto -h_x(T-t)$. Hence, it can be used to distinguish the two degenerate optimal protocols. The trajectory of the mixed state after tracing out one qubit on the Bloch sphere is shown in \href{https://mgbukov.github.io/movies/2B_paper/Movie-1a.mp4}{Movie 1a} and \href{https://mgbukov.github.io/movies/2B_paper/Movie-1b.mp4}{Movie 1b}. This is an indication that the control phases depend strongly on the cost function used to set up the optimisation problem.

Similar behavior is observed in the expectation values of other observables, see Fig.~\ref{fig:Sent_split}(b). While both optimal protocols lead to states which are globally equally far away from the target, \emph{locally} one is closer than the other. Intuitively, one anticipates this to be the low-entangled state, since the final state is also weakly entangled. The expectation values of the local operators $\langle\psi(T)| S^x_j|\psi(T)\rangle$ and $\langle\psi(T)| S^z_j|\psi(T)\rangle$ actually show the opposite behavior. To reconcile these observations, we  compute the local Uhlmann fidelity, 
\begin{equation}
f_h(T) = \left(\mathrm{tr}\sqrt{ \sqrt{\rho(T)}\rho_\ast\sqrt{\rho(T)} }\right)^2,
\end{equation}
where $\rho(T)$ and $\rho_\ast$ are the reduced density matrices of the evolved and target states.  The Uhlmann fidelity measures how distinguishable the final and target states are, if we perform the optimal local measurement that distinguishes the target from the evolved state. As expected, this criterion shows that the high-entangled state is further away from the target state than the low-entangled state. However, for some observables, the expectation value of an operator in the high-entangled state reflects more accurately its target-state value compared to its expectation value in the low-entangled state. Thus, for all practical purposes, whether the high or low-entangled states are closer to the target ground state strongly depends on the actual quantity of interest.

With the advent of recent advances in experimental physics, it is within the scope of highly-controllable present-day experiments to measure the entanglement entropy~\cite{walborn_06,kang_12,zhou_14,islam_15,cheng_16}. This is not as complicated in a two-qubit system, since the entanglement shared between two qubits can be inferred directly from a measurement of the local magnetization. In this respect, the bifurcation of entanglement and observables in the symmetry-broken phase close to optimality serves as a smoking gun to directly probe the physics of this correlated quantum control phase.

\section{\label{sec:variational}Effective Variational Theory for Symmetry Breaking and High-Fidelity Protocols}

The usefulness of the optimal protocols depends on their robustness to small perturbations. It has recently been shown that the optimal protocol can be unstable in the glassy phase of quantum state preparation in a nonintegrable system with many coupled qubits~\cite{bukov_17RL}. Nevertheless, we demonstrate that there exist simple, nearly optimal but robust solutions even in the symmetry-broken phase. To capture the properties of the good protocols lying low in the infidelity landscape, we consider a family of three-pulse protocols, the pulse lengths of which define variational parameters. This family allows for symmetry breaking, yet the latter is not enforced. By optimising the best achievable fidelity within this three-pulse variational manifold, we can capture the overconstrained-to-correlated critical point $T_c$ and the spontaneous symmetry-breaking point $T_\mathrm{sb}$. Moreover, this ansatz likely yields the optimal protocol for the entire range $T\leq T_\mathrm{sb}$. Yet, it is inferior to SD for $T>T_\mathrm{sb}$ and, thus, fails to capture the QSL point $T_\mathrm{QSL}$, at least within the short protocol durations of interest, presumably due to the glassy character of the landscape in the symmetry-broken phase. Quite generally, one can think of such a variational ansatz as an affective mean-field theory for the quantum control optimisation problem.

\begin{figure*}[t!]
	\includegraphics[width=1.0\textwidth]{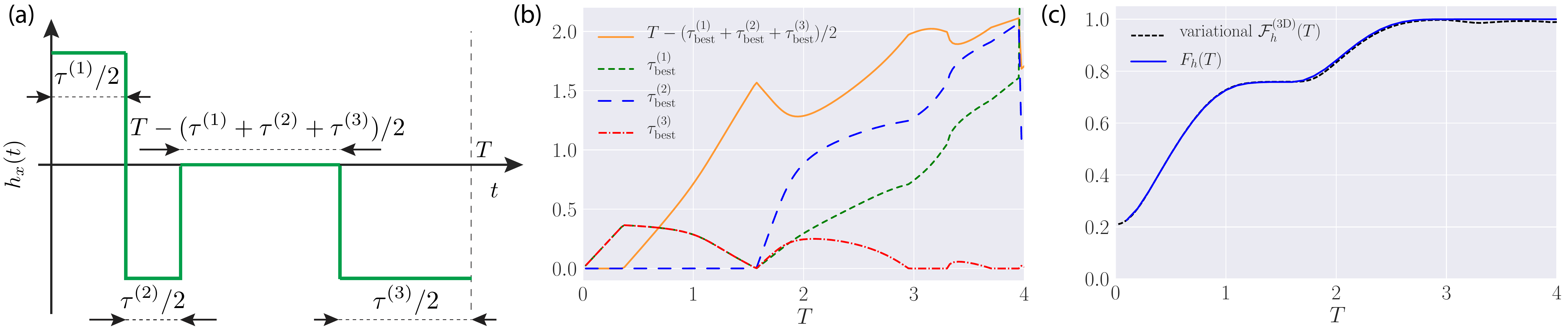}
	\caption{\label{fig:var_theory} Variational theory for the symmetrically-coupled two-qubit system. (a) generic form of the family of variational protocols, with pulse durations $\tau^{(i)}$, which allow for symmetry breaking, to be chosen by optimising the fidelity (see text). Optimal pulse durations (b) against the total protocol time $T$ feature kinks at the durations corresponding to the control phase transitions. (c) The optimal variational fidelity (dashed line), compared to the best fidelity obtained using SD. Time is in units of the inverse qubit interaction strength. }
\end{figure*}

Inspired by the behaviour displayed by the sample-average protocols at all $T$, see Fig.~\ref{fig:protocol_averaged}, we consider the four-pulse sequence shown in Fig.~\ref{fig:var_theory}a. Define the variational infidelity landscape as
\begin{widetext}
	\begin{eqnarray}
	\mathcal{I}_h(\tau^{(i)};T) &=& 1-\mathcal{F}_h(\tau^{(i)};T),\nonumber\\ 
	\mathcal{F}^\mathrm{(3D)}_h(\tau^{(i)};T) &=&\left|\langle\psi_\ast|\mathrm{e}^{-i\frac{\tau^{(3)}}{2} H[h_x=-4]}\mathrm{e}^{-i\left(T-\frac{\tau^{(1)}+\tau^{(2)}+\tau^{(3)}}{2}\right) H[h_x=0]}\mathrm{e}^{-i\frac{\tau^{(2)}}{2} H[h_x=-4]}\mathrm{e}^{-i\frac{\tau^{(1)}}{2} H[h_x=4]} |\psi_i\rangle\right|^2,
	\label{eq:var_ansatz}
	\end{eqnarray}
\end{widetext}
which is a function of the three pulse lengths $\tau^{(i)}$, and depends parametrically on the total protocol duration $T$. Here $H[h_x]$ is the two-qubit Hamiltonian~\eqref{eq:H_2B}. Thus, this defines a variational problem
\begin{equation}
\partial_{\tau^{(i)}} \mathcal{I}(\tau^{(i)};T) = 0,\qquad 0\leq \tau^{(i)}\leq T
\end{equation}
which is only three dimensional, and can be solved numerically to determine the optimal pulse lengths $\tau^{(i)}_\mathrm{best}$. 

\begin{figure*}[t!]
	\includegraphics[width=1.0\textwidth]{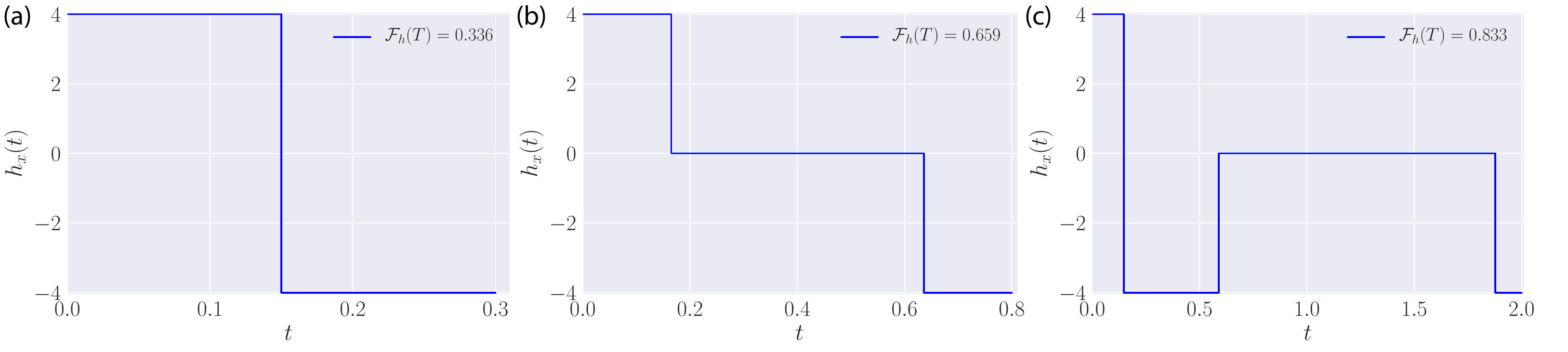}
	\caption{\label{fig:var_protocols}Variational protocols for $T=0.3$ (a), $T=0.8$ (b) and $T=2.0$ (c). At $T=2.0$, the best variational protocol breaks the $\mathcal{Z}_2$ symmetry: $h(t)\neq -h(T-t)$. Time is in units of the inverse qubit interaction strength.}
\end{figure*}

Figure~\ref{fig:var_protocols} shows the variational protocols which minimise $\mathcal{I}_h(T)$ in the different phases. For $T<T_c$, we find $\tau^{(1)}_\mathrm{best}=\tau^{(3)}_\mathrm{best}=T$ and $\tau^{(2)}_\mathrm{best}=0$. Note that there is an ambiguity in which one of the two variables $\tau^{(2)}_\mathrm{best}$ and $\tau^{(3)}_\mathrm{best}$ to keep finite here, as can be seen from the definition of the variational protocol, cf.~Fig.~\ref{fig:var_theory}. However, only one of them can be non-zero for $T<T_c$. As a result, the variational protocol features a single bang at half the protocol duration, see Fig.~\ref{fig:var_protocols}a. At the critical point $T_c$, $\tau^{(1)}_\mathrm{best}=\tau^{(3)}_\mathrm{best}<T$ and another pulse appears in the middle of the protocol during which the control field $h_x=0$, see Fig.~\ref{fig:var_protocols}b. Notice that in the overconstrained and symmetric correlated phases the variational solution is symmetric $h_x(t)=-h_x(T-t)$, although no symmetry has been imposed explicitly, in agreement with the observations from the main text. Beyond the symmetry-breaking critical point $T_\mathrm{sb}$, the infidelity is minimised for a finite pulse length $\tau^{(2)}_\mathrm{best}>0$. As a result, the variational protocol is symmetry-broken and degenerate, see Fig.~\ref{fig:var_protocols}c. Note that, in the symmetry-broken glassy phase $\tau^{(1)}_\mathrm{best}\neq \tau^{(3)}_\mathrm{best}$. 

The variational fidelity $\mathcal{F}_h(\tau^{(i)};T)$ corresponding to the best protocols is shown in Fig.~\ref{fig:var_theory}. A comparison with the best numerical fidelity, c.f.~\ref{fig:var_theory}c, reveals that in the symmetric phases, $T<T_\mathrm{sb}$, the simple variational ansatz in fact captures the global minimum of the infidelity landscape, while it is clearly suboptimal in the symmetry-broken phase. Nevertheless, its performance rivals that of the optimal solution in the entire protocol duration range of interest. The trajectory of the mixed state after tracing out one qubit on the Bloch sphere for the variational protocol and its symmetry-related partner is shown in \href{https://mgbukov.github.io/movies/2B_paper/Movie-2a.mp4}{Movie 2a} and \href{https://mgbukov.github.io/movies/2B_paper/Movie-2b.mp4}{Movie 2b} for $T=2.0$ [to be compared with the \href{https://mgbukov.github.io/2B_movies}{solution obtained using SD}].

The overconstrained and unbroken correlated/glassy phases share many [and probably all] properties of their single-qubit counterparts. Hence, the critical point $T_c$, as well as the structure of the optimal protocols, can be understood in terms of a renormalised single-qubit variational theory.

\section{\label{sec:outro}Conclusions}

State preparation in the symmetrically-coupled two-qubit problem exhibits a rich control phase diagram. Apart from an overconstrained, correlated (glassy), and controllable phases, the optimal solution is double degenerate in a broad region of protocol durations just before the quantum speed limit as a consequence of breaking a discrete symmetry in the quantum control landscape. Being a property of the control landscape, these phase transitions are present in any optimization algorithm with local (in time) flip updates, as we verified explicitly using SD, Reinforcement Learning and GRAPE. We also verified that all control critical points and phases are not sensitive to the family of bang-bang protocols we used, by using GRAPE to study this optimization problem in an experimentally more relevant set of continuous protocols $|h_x(t)|\leq 4$.

The results of this paper show the importance of the cost function in quantum optimal control. The final states in the symmetry-broken phase are degenerate regarding their global distance to the target state, but one locally resembles the target better than the other. The symmetry breaking moreover highlights the potential importance of singular regions in quantum control problems, where straightforward application of Pontryagin's principle fails. Indeed, all variational protocols deviate from bang-bang over finite time intervals and constitute so called bang-singular control. 

For the present model, the symmetry broken phase is absent for all quantum spin chains with $L\neq 2$~\cite{bukov_17RL}, or when the objective is extended to prepare all three eigenstates of the target Hamiltonian with the same protocol. Hence, the mechanisms for the appearance of the symmetry breaking in the control landscape remains an open problem for future investigation.

\begin{acknowledgments}
	MB acknowledges support from the Emergent Phenomena in Quantum Systems initiative of the Gordon and Betty Moore Foundation (\href{https://www.moore.org/initiative-strategy-detail?initiativeId=emergent-phenomena-in-quantum-systems}{EPiQS}) and ERC synergy grant \href{http://www.uquam.eu/}{UQUAM}. AD is supported by a NSERC PGS D. DS acknowledges support from the FWO as post-doctoral fellow of the Research Foundation -- Flanders and CMTV. MB, PW and AP were supported by NSF DMR-1506340 and AFOSR FA9550-16-1-0334. AD and PM acknowledge support from the Simons Foundation through the MMLS Fellow program. We used \href{http://weinbe58.github.io/QuSpin/}{QuSpin} for simulating the dynamics of the qubit system~\cite{weinberg_17,quspin2}. The authors are pleased to acknowledge that the computational work reported on in this paper was performed on the Shared Computing Cluster which is administered by \href{https://www.bu.edu/tech/support/research/}{Boston University's Research Computing Services}. The authors also acknowledge the Research Computing Services group for providing consulting support which has contributed to the results reported within this paper.
\end{acknowledgments}

\bibliographystyle{apsrev4-1}
\bibliography{control_RL.bib}

\appendix

\newpage

\begin{widetext}
	
	\section{\label{app:scaling}Scaling Analysis of the Control Critical Points}

	In the main text, we discussed the existence of various phase transitions in the control landscape of the state preparation problem in the symmetrically-coupled two-qubit system. We also explained that these transitions occur in the low infidelity (a.k.a.~``energy") manifold of an effective classical Ising spin model $\mathcal{H}_\mathrm{eff}(T)$, describing the quantum state preparation control problem, featuring nonlocal multi-body all-to-all interactions~\cite{day_17}. Here we present the finite-size scaling curves for the important quantities which reveal the control phase transitions. Notice that, even though our system has only three quantum levels, the effective underlying spin model $\mathcal{H}_\mathrm{eff}(T)$ describes the physics of a many-body system with many degrees of freedom. Recall that the lattice constant for $\mathcal{H}_\mathrm{eff}(T)$ is set by the protocol time step $\delta t$. Hence, the finite-size scaling should be done in the continuum limit $\delta t\to 0$ with the total number of bang-bang steps $T/\delta t=N_T=const$.
	
	\begin{figure}[h!]
		\includegraphics[width=1.0\columnwidth]{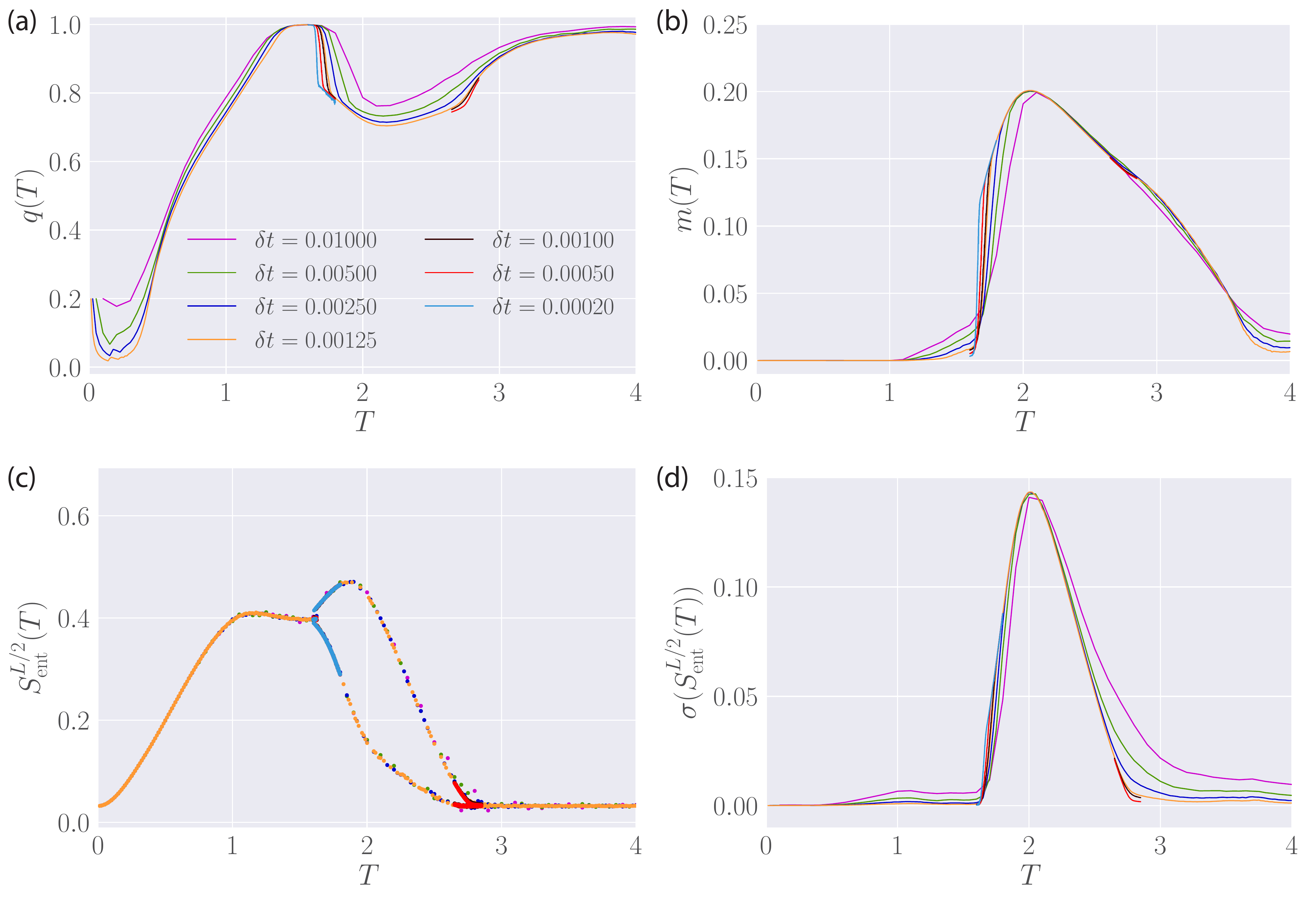}
		\caption{\label{fig:scaling} Protocol time step ($\delta t$) scaling of the quantities used to detect the control phase transitions: (a) Edwards-Anderson-like order parameter, (b) magnetisation order parameter, (c) entanglement entropy of the (nearly) optimal best encountered protocol, and (d) standard deviation of of the entanglement entropy over the sample of local infidelity minima. The sample size of local infidelity minima contains $N_\mathrm{real}=10^3$ independent realisations obtained using Stochastic Descent (SD). The time step decreases from the topmost to the lowermost curve, and the transitions become sharper with decreasing $\delta t$. Time is in units of the inverse qubit interaction strength.}
	\end{figure}
	
	Figure~\ref{fig:scaling}(a) shows the finite-size scaling of the main quantities of interest. Let us focus first on the order parameter -- the Edwards-Anderson-like correlator $q(T)$. Observe that the overconstrained-to-symmetric glass critical point $T_c$, see Fig.~\ref{fig:phase_diag_2B} (main text) emerges clearly in the limit $\delta t\to 0$. We mention that this transition is present also in the single qubit limit $J\to 0$~\cite{bukov_17RL}, where an exact expression can be obtained. Therefore, we expect that, while for $J>0$ the exact expression for $T_c$ is modified, the underlying physics remains the same. The symmetry-breaking critical point $T_\mathrm{sb}\approx 1.55$ is discontinuous, since the correlator $q(T)$ exhibits a sharp jump across it. Indeed, Fig.~\ref{fig:scaling}(a) shows the emergence of a jump for $T\to T_\mathrm{sb}^-$ where all protocols are uncorrelated and $q(T_\mathrm{sb})=1$, as opposed to the symmetry-broken phase with correlated local infidelity minima for $T\to T_\mathrm{sb}^+$. At this point, the optimal protocol breaks the symmetry of the problem and becomes doubly degenerate. The controllability critical point appears at $T_\mathrm{QSL}\approx 2.8$ and comes with a kink in the order parameter $q(T)$. Interestingly, it takes an order of magnitude smaller protocol step size $\delta t$ to resolve it, compared to the $J=0$ case.
	
	Fig~~\ref{fig:scaling}(b) shows the finite-size scaling of the magnetisation order parameter $m(T)$. Once again, a sharp jump becomes visible at the symmetry-breaking point $T_\mathrm{sb}$, supporting the discontinuous character of this control phase transition, at least in the family of bang-bang protocols. Interestingly, at $T\approx 3.5$ in the controllable phase, the magnetisation curves cross again. It is currently an open question whether this is associated with yet another continuous symmetry-restoration transition in the limit $\delta t\to 0$ inside the controllable phase. 
	
	Fig~\ref{fig:scaling}(c-d) shows the critical scaling of the entanglement entropy $S_\mathrm{ent}(T)$, associated with the optimal protocol, and its standard deviation computed over the sample of infidelity minima.

	\section{\label{app:Tsb_derivation}Determining the Symmetry-Breaking Critical Point}
	
	In this section, we determine the dependence of the symmetry-breaking critical point on the model parameters. While we do not have a complete theory for this transition, it is still possible to derive an equation for $T_\mathrm{sb}$ as follows. Let us draw the attention of the reader to the following important observations:
	\begin{itemize}
		\item[(i)] motivated by Fig.~\eqref{fig:protocol_averaged}(c) and general symmetry arguments (see main text), we make the ansatz that the optimal protocol at $T=T_\mathrm{sb}$ vanishes identically: $h_x(t)\equiv 0$,
		\item[(ii)] since the variational ansatz of Eq.~\eqref{eq:var_ansatz} in fact produces the \emph{optimal} protocol for $T<T_\mathrm{sb}$, and breaks precisely at the symmetry-breaking critical point, we can extract $T_\mathrm{sb}$ as the largest protocol duration the ansatz $h_x(t)\equiv 0$ is valid for.
	\end{itemize}
	Combining the two points, we have to maximise the following fidelity
	\begin{equation}
	\mathcal{F}_\mathrm{sb}(T)=\left|\langle\psi_\ast|\mathrm{e}^{-iT[-2JS_1^zS_2^z-g(S^z_1+S^z_2)]} |\psi_i\rangle\right|^2 
	\end{equation}
	which results in a transcendental equation for $T_\mathrm{sb}$:
	\begin{equation}
	-4 b g\sin(2 g T_\mathrm{sb}) + 2 a [b (g - J)\sin([g - J] T_\mathrm{sb}) + (g + J) \sin([g + J] T_\mathrm{sb})]=0,
	\label{eq:F_Tsb}
	\end{equation}
	where
	\begin{eqnarray*}
		a &=& \frac{(6 g + 2 J + s)^2}{18 h_{x,i}^2}\nonumber\\
		b &=& \left(1+\frac{(4 J - s) (6 g + 2 J + s)}{18 h_{x,i}^2}\right)^2\nonumber\\
		s &=& -4 \sqrt{3 (h_{x,i}^2 + g^2) + J^2}\sin\left(\frac{\pi}{6} + \frac{1}{3}\arccos\left[
		\frac{J}{2} \left(\frac{1}{3 (h_{x,i}^2 + g^2)+ J^2} \right)^{3/2} (9 h_{x,i}^2 + 2 (J^2-9 g^2)) \right]\right) \nonumber
	\end{eqnarray*}
	Here $h_{x,i}=-2=-h_{x,\ast}$ denotes the value of the $x$-field corresponding to the initial and target states, respectively, which ensures the existence of the $\mathbb{Z}_2$ symmetry in protocol space. 
	
	Notice that for $J=g$, Eq.~\eqref{eq:F_Tsb} reduces to $\sin(2 g T_\mathrm{sb})=0$, which for $g=1$ has the unique solution $T_\mathrm{sb}=\pi/2$ with fidelity $\mathcal{F}_\mathrm{sb}(T_\mathrm{sb})\approx 0.759252$. This result is independent of the values for $J$ and $h_{x,i}$, and in excellent agreement with the numerical simulations.
	
	\section{\label{app:var_control}Variational Proof of Controllability}
	\begin{figure}[t!]
		\includegraphics[width=1.0\columnwidth]{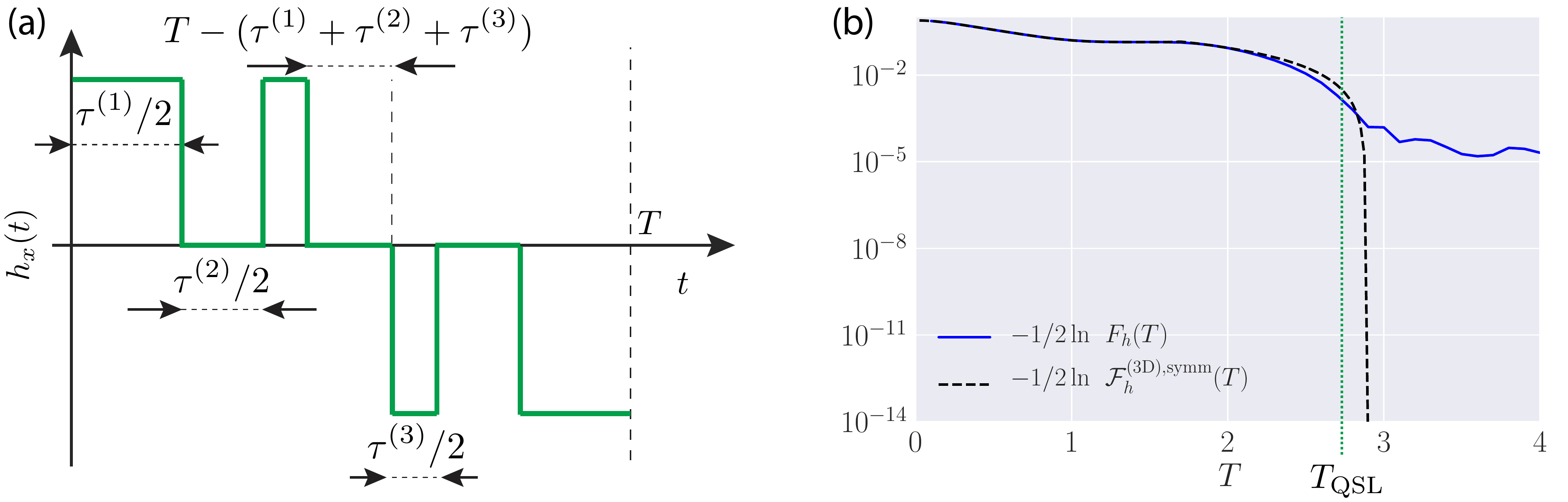}
		\caption{\label{fig:var_control} (a) Variational proof of controllability of the system, using a simple 3D symmetric ansatz (black dashed line). (b) The optimal protocol within this variational family reaches unit fidelity, which is marked by a vertical asymptote in the logarithmic variational fidelity. The green dotted line shows the position of the true QSL for the system. For comparison, the fidelity obtained using SD is also shown (solid blue line). Time is in units of the inverse qubit interaction strength.}
	\end{figure}
	
	We can give a constructive proof for the controllability of the symmetrically-coupled two-qubit system, using a variational ansatz as follows. Similar to the discussion on the effective variational theories above, where we made an ansatz allowing for symmetry breaking of the variational protocol, we make the following three-pulse \emph{symmetric} variational ansatz. 
	\begin{eqnarray}
	\mathcal{F}^\mathrm{(3D),symm}_h(\tau^{(i)};T) &=&\bigg|\langle\psi_\ast|
	\mathrm{e}^{-i\frac{\tau^{(1)}}{2} H[h_x=-4]}
	\mathrm{e}^{-i\frac{\tau^{(2)}}{2} H[h_x=0]}
	\mathrm{e}^{-i\frac{\tau^{(3)}}{2} H[h_x=-4]}\times\nonumber\\
	&&\phantom{\bigg|\langle\psi_\ast|}\times
	\mathrm{e}^{-i\left(T-\tau^{(1)}-\tau^{(2)}-\tau^{(3)}\right) H[h_x=0]}\nonumber\\
	&&\phantom{\bigg|\langle\psi_\ast|}\times
	\mathrm{e}^{-i\frac{\tau^{(3)}}{2} H[h_x=4]}
	\mathrm{e}^{-i\frac{\tau^{(2)}}{2} H[h_x=0]}
	\mathrm{e}^{-i\frac{\tau^{(1)}}{2} H[h_x=4]}
	|\psi_i\rangle\bigg|^2,
	\label{eq:var_ansatz_control}
	\end{eqnarray}
	where the variables $\tau^{(i)}$ are determined by solving the associated optimisation problem. Since the resulting expressions are rather cumbersome, we refrain from showing them explicitly. This sequence is shown schematically in Fig.~\ref{fig:var_control}a. Since the symmetry of the protocol is hard-coded into it, the ansatz~\eqref{eq:var_ansatz_control} cannot capture the optimal protocol for the symmetry broken phase by construction.
	
	Nevertheless, this simple ansatz demonstrates that the system is indeed controllable, as the optimal variational protocol reaches unit fidelity at $T\approx 2.907$, although a bit after the true quantum speed limit $T_\mathrm{QSL}\approx 2.775$, independently estimated within the precision of the numerical algorithms Stochastic Descent and GRAPE. Figure.~\ref{fig:var_control}b (dashed black line) shows the logarithmic optimal fidelity within this 3D symmetric ansatz. The presence of the vertical asymptote is a clean numerical proof for the controllability of the system. As the ansatz is suboptimal, this happens for a protocol duration $T\approx 2.907>2.775\approx T_\mathrm{QSL}$ greater than the true QSL (green dotted line).
	
\end{widetext}

\end{document}